\documentstyle[12pt]{article}
\makeatletter

\@addtoreset{equation}{section}
\makeatother
\def\abstract#1{{\centerline{\bg Abstract}} \vskip 3mm \par #1}
\def\cy{Calabi-Yau}
\def\cym{Calabi-Yau manifold}
\def\lg{Landau-Ginzburg}

\def\tg{\tilde{\theta_{i}}^{g}}

\def\thd{\tilde{\theta_{i}}^{h^{\prime}}}
\def\al{\alpha}

\def\vp{\varphi}
\def\inbar{\vrule height1.5ex width.4pt depth0pt} %\font\ZZsf=cmss12

\def\ZZ{\relax{\sf Z\kern-.4em \sf Z}}  \def\IR{\relax{\rm  
I\kern-.18em R}}
\def\IN{\relax{\rm I\kern-.18em N}} \def\IP{\relax{\rm I\kern-.18em  
P}}
\def\IQ{\relax\,\hbox{$\inbar\kern-.3em{\rm Q}$}}
\def\la{\lambda}
\def\IC{\hbox{\,$\inbar\kern-.3em{\rm C}$}}

\def\({\lbrack}
\def\){\rbrack}

\def\ketc#1{{\left| #1\right\rangle}_{\rm (c,c)}}
\def\keta#1{{\left| #1\right\rangle}_{\rm (a,c)}}

\begin{document}
\baselineskip=6mm
\begin{flushright}
{hep-th/9612239} \\
{OU-HET 255} \\
%{hep-th/9612239} \\
{December 1996}
\end{flushright}
\vskip 1.0cm
\centerline{\LARGE {\bf Mirror Symmetry and the Web of }}
\vskip 0.5cm
\centerline{\LARGE {\bf \lg\ String Vacua }}
%\vskip 1cm
%\centerline{\LARGE {\bf and}}
%\vskip 1cm
%\centerline{\LARGE {\bf Landau-Ginzburg Orbifolds}}
\vskip 1.0cm
\centerline{\large Hitoshi \ Sato}
\vskip 1cm
\centerline{\it Department of Physics, Faculty of Science, Osaka  
University}
\centerline{\it Toyonaka, Osaka 560, Japan} 
\centerline{email address : sato@funpth.phys.sci.osaka-u.ac.jp}
%\maketitle
\vskip 1.5cm
%\maketitle
\centerline{\large {\bf ABSTRACT} }

\vskip 0.5cm

We  present some mathematical aspects of \lg\ string vacua
in terms of toric geometry.
The one-to-one correspondence between toric divisors and 
some of $(-1,1)$ states in \lg\ model is presented
for superpotentials of typical types.
%for some classes of superpotentials.
The \lg\ interpretation of non-toric divisors is also
presented.
Using this interpretation, we propose a method to solve
the so-called "twisted sector problem" by orbifold construction.
Moreover,this construction shows that 
the moduli spaces of the original
\lg\ string vacua and their orbifolds are connected.
By considering the mirror map of \lg\ models,
we obtain the relation between Mori vectors 
and the twist operators of our orbifoldization.
This consideration enables us to argue 
the embedding of the Seiberg-Witten curve
in the defining equation of the \cym s
on which the type II string gets compactified.
%in terms of the orbifold twists.
Related topics concerning the \cy\ fourfolds
and the extremal transition are discussed.

\thispagestyle{empty}

%%%%%%%%%%%%%%%%%%%%%%%%%%%%%%%%%%%%%%%%%%%%%%%%%%%%%%%%%%%%%%%%%%%%% 
%%%%%%
%article
%%%%%%%%%%%%%%%%%%%%%%%%%%%%%%%%%%%%%%%%%%%%%%%%%%%%%%%%%%%%%%%%%%%%% 
%%%%%%
\clearpage

\pagenumbering{arabic}

\section{Introduction}

Strings on \cym s are promising candidates 
for unified theory.
Their physics heavily depends on the topological
properties of \cym s.
Recently it is believed that the toric geometry 
gives us the most useful technique
to study \cym s \cite{vb,hkty1}.

It is also known that \lg\ models 
\cite{iv} can describe the strings
on \cym s.
They give us a powerful method for calculating
the (anti-)chiral rings \cite{lvw}
which can be identified with the cohomology rings
on the corresponding \cym s,
i.e.$(p,q)$ states of \lg\ model
can be identified with 
$(3-p,q)$ forms on a \cym .

In this paper, we study the relation between \lg\ models
and toric geometry.
First, we find the one-to-one identification
between a toric divisor and a $(-1,1)$ state
which is a ground state in $j^{-l}$ twisted sector,
where $j$ is the $U(1)$ twist of our \lg\ model.
In the case of Fermat type potential
this identification was already given
by the author 
\cite{sa3}.
We try to extend our results to the potentials of 
other types, i.e. chain and loop potentials.

In \cite{sa3} 
it was not possible to identify
the $(-1,1)$ states written in the form
$\prod_{i}{{X_{i}}^{l_{i}}} \keta {j^{-l}}$
with toric data.
In this paper, however,
 we have succeeded in  the toric interpretation
of these states.
Futhermore we can show 
that these states correspond to the 
so-called "twisted sector problem" in \cite{hkty1}.
Roughly speaking,
this problem implies that 
toric geometry fails to 
give us the techniques for describing 
the  whole K\"{a}hler moduli space
and the whole complex structure moduli space.
It was pointed out that under certain circumstances
we can solve this problem by deforming a polyhedron
obtained from a defining equation of a \cym\ \cite{klemm} .
We find that in the \lg\ context this construction 
corresponds to the orbifold construction of \cite{ls}.
This understanding enables us to propose a useful method
to solve the twisted sector problem.
Furthermore we will show that the Mori vectors,
%which areFurthermore, we will show that the Mori vectors,
which are needed to study the moduli space
of \cym s, 
can be obtained from the relation of 
the twist operators of our orbifoldization.
This orbifold construction tells us
that the moduli spaces of the original \lg\ string vacua
and their orbifolds are connected.
This structure can be shown in terms of 
our correspondence
between \lg\ models and toric geometry.
%It is well known that
%to study the moduli space of \cym s,
%we need to obtain Mori cone \cite{hkty1}.

This paper is organized as follows.
In section 2 we briefly review the construction
of \cym s in terms of toric geometry.
The twisted sector problem is explained here.
The \lg\ models and their mirror symmetry 
are given in section 3.
To study mirror symmetry
the useful ingredients are 
the phase symmetries of the superpotentials
for the \lg\ models.
This procedure is explained in this section.
In terms of the phase symmetries the correspondence between
\lg\ models and toric geometry is presented.

Section 4 is devoted to
the orbifold construction
and its toric aspects.
In the previous section
we have presented the one-to-one identification
between the $(-1,1)$ states and the toric data.
Using this identification,
a useful method to 
resolve the twisted sector problem is proposed.
This mehtod is based on the orbifold construction.
For \lg\ models of our interest,
this orbifold construction 
requires the deformation of the superpotential.
This deformation should correspond to 
the deformation of toric data.
As a result the moduli spaces of the original
and the orbifoldized theories are connected.
We will show the useful formula
to obtain the appropriate twist operators for this orbifoldization
in terms of the \lg\ analysis.
The relation between the phase symmetries and the Mori vectors
are also discussed.
Typical examples are studied in detail.

In section 5 we discuss another application of 
our toric interpretation of \lg\ models
in the context of $K3$-fibred \cym s and 
the Seiberg-Witten geometry.
An interesting example is given
from which
we can obtain a quantum field theory
at the orbifold point in the moduli space.
Finally, in section 6
some speculative remarks are given.

\section{General construction of \cy\ hypersurfaces
in toric varietes}

Let us briefly review a method of toric geometry
\cite{vb,hkty1}.
In toric geometry,a pair of   
$ (\Delta , \Delta^*) $ gives a \cym ,
where $ \Delta $ is a (Newton) polyhedron corresponding to
monomials and $ \Delta^* $ is a dual (or polar) polyhedron
describing the resolution of singularities.
A point on a one- or two-dimensional face
of $\Delta^*$ corresponds to a $(1,1)$ form
coming from resolution. 

First, we consider the Fermat type quasi-homogeneous polynomial
in the weighted complex projective space  
$WCP^{4}_{(w_{1},w_{2},w_{3},w_{4},w_{5})}[d]$,
where $w_{i}$ are weights and 
the degree $ d = \sum_{i = 1}^{5}{w_{i}}$.
The corresponding \cy\ hypersurface consists of
monomials $z_i^{d/w_i}\,\,(i=1,\cdots,5)$.
We assume  $w_{1} = 1$, 
so that the polyhedron obtained from the hypersurface
is $4$-dimensional.
The associated $4$-dimensional integral convex polyhedron
$\Delta(w)$ 
is the convex hull of the integral vectors $m$ of the exponents
of all quasi-homogeneous monomials 
of degree $d$
shifted by $(-1,\ldots,-1)$, i.e.
$\prod_{i=1}^{5}{z_{i}^{m_{i}+1}}$: 
\begin{equation}
\Delta(w):=
\{(m_1,\ldots,m_{5}) \in
\IR^{5}|\sum_{i=1}^{5} w_i m_i=0,m_i\geq-1\} .
\end{equation}
This implies that only the origin is 
the point in the interior of $\Delta$.
Its dual polyhedron is defined by
\begin{equation}
\Delta^*=\{\;(x_1,\dots,x_4) \;\vert\;
\sum_{i=1}^4 x_i y_i \geq -1 \;{\rm for \; all \; }
(y_1,\dots,y_4)\in\Delta\;\}.
\end{equation}
In our case it is known that 
$ (\Delta , \Delta^*) $
is a reflexive pair.
An $l$-dimensional face 
$\Theta \subset \Delta$ can be represented by specifying
its vertices
${\rm v}_{i_1},\cdots,{\rm v}_{i_{k}}$. 
Then the dual face $\Theta^*$
is a $(4-l-1)$-dimensional face of $\Delta^*$
and defined by
\begin{equation}
\label{dualface}
\Theta^*=\{ x\in \Delta^*\; 
\vert \; (x,{\rm v}_{i_1})=\cdots=(x,{\rm
v}_{i_k})=-1 \},
\end{equation}
where $(*,*)$ is the ordinary inner product.

For models of Fermat type,
we then always obtain as
vertices of $\Delta(w)$
\begin{eqnarray}
\label{fermatpolyhedron}
&\nu_1&=(-1,-1,-1,-1),
\nu_2 \! \! \! =(d/w_2-1, -1, -1, -1),  \  
\nu_3=(-1, d/w_3-1, -1, -1), \nonumber \\ 
&\nu_4&=(-1, -1, d/w_4-1, -1), \ 
\nu_5 \! \! \! =(-1, -1, -1, d/w_5-1),
\end{eqnarray}
and for the vertices of the dual polyhedron 
$\Delta^*(w)$ one finds
\begin{eqnarray}
&\nu_1^*&=(-w_2,-w_3,-w_4,-w_5), \nonumber \\
&\nu_2^*&=(1,0,0,0),  \quad 
\nu_3^*=(0,1,0,0), \quad
\nu_4^*=(0,0,1,0), \quad
\nu_5^*=(0,0,0,1).
\end{eqnarray}

Batyrev showed the following useful formula 
for non-trivial cohomologies \cite{vb};
\begin{equation}
\label{vb11}
h^{1,1} = 
l(\Delta^{*}) - 5 
- \sum_{\bf{codimension} \; {\Theta^{*}} = 1}
{l^{'}({\Theta^{*}})}
+\sum_{\bf{codimension} \; \Theta^{*} = 2, 
\Theta^{*} \in \Delta^{*}}
{l^{'}(\Theta^{*})
l^{'}(\Theta)}
,
\end{equation}
\begin{equation}
\label{vb21}
h^{2,1} = 
l(\Delta) - 5 
- \sum_{\bf{codimension} \; {\Theta} = 1}
{l^{'}({\Theta})}
+ \sum_{\bf{codimension} \; \Theta = 2, 
\Theta \in \Delta}
{l^{'}(\Theta)
l^{'}(\Theta^{*})}
,
\end{equation}
where 
${l({\Delta})}$ ($l(\Delta^{*})$)
denotes the number of integral points
in ${\Delta}$ (${\Delta}^{*}$).
The symbol 
${l^{'}({\Theta})}$ ($l^{'}(\Theta^{*})$)
denotes the number of integral points
in the interior of ${\Theta}$ (${\Theta}^{*}$).

%If the last terms of (\ref{vb11}) and (\ref{vb21})
%do not vanish,
%there are `twisted sector problem'.
The non-zero contributions of 
the last terms of (\ref{vb11}) and (\ref{vb21}) 
are called "twisted sector problem" \cite{hkty1}
If this problem occurs,
toric geometry is not sufficient to 
give us the techniques for describing 
the  whole K\"{a}hler moduli space
and the whole complex structure moduli space.
In section 4,
we will explain this problem more in depth
and propose a method to circumvent 
this undesirable situation.

%In the case of 
For the non-Fermat polynomials 
which always contain the another type of 
monomials ${z_{i}}^{a_{i}}z_{j}$,
it was shown \cite{klemm,ca1} that toric geometry works 
in a similar manner.
For example, if the defining equation contains 
the monomial ${z_{4}}^{a_{4}}z_{5}$
in stead of $z_{4}^{d/w_{4}}$,
we have to consider the vertex 
$(-1,-1,a_{4}-1,0)$ instead of $\nu_{4}$
in (\ref{fermatpolyhedron}).

The defining equation for
a mirror manifold 
is obtained as 
the zero locus of the Laurent polynomial 
\begin{equation}
p^{'} = \sum_{i}{a_{i}{{y_{i}}^{{\nu_{i}}^{*}}}}
\end{equation}
where $y_{i}$ are 
coordinates of a canonical torus $({\rm C^{*}})^{4}$
and the coefficients $a_{i}$ are parameters 
characterizing the complex structure of the mirror manifold.
If one wants to get the hypersurface as a 
quasihomogeneous polynomial,
one should specify an \'{e}tale map \cite{klemm};
\begin{equation}
\label{etale}
y_{i} = \frac{\bar{m_{i}}}
{\bar{X_{1}}\bar{X_{2}}\bar{X_{3}}\bar{X_{4}}\bar{X_{5}}},
\end{equation}
where $\bar{X_{i}}$ are the fields 
and $\bar{m_{i}}$ is the appropriate monomial 
of the transposed polynomial \cite{bh}
which will be explained in section 3.1 
in the context of the \lg\ superpotentials.

\section{Toric geometrical aspects of \lg\ string vacua}

\subsection{\lg\ models and their mirrors}

A \lg\ model is characterized by a superpotential
$W(X_{i})$ where $X_{i}$ are $N = 2$ chiral superfields.
%In this paper, we will restrict our attention to 
%the superpotential of a form
%$W{(X_{i})} = 
%X_{1}^{a_{1}}+X_{2}^{a_{2}}+X_{3}^{a_{3}}
%+X_{4}^{a_{4}}+X_{5}^{a_{5}},$
%which corresponds to the Fermat type hypersurface
%in $WCP^{4}$.
The \lg\ orbifolds 
\cite{iv} are obtained by
quotienting with an Abelian symmetry group $G$ of 
$W{(X_{i})}$,
whose element $g$ acts as an $N \times N$ diagonal matrix,
$g: X_{i} \rightarrow e^{2 \pi i {\tg}}X_{i}$.
Here $0 \leq \tg < 1$.
Of course the $U(1)$ twist \ 
$j: X_{i} \rightarrow e^{2 \pi i {q_{i}}}X_{i}$ \ 
generates the symmetry group of
$W{(X_{i})}$,
where $q_{i} = {w_{i} \over d}$,\ \ 
 $W(\la^{w_{i}} X_{i}) = \la^{d}W(X_{i})$ \ and
$\la \in \IC^{\ast}$.
In this paper, we further require that 
$ w_{1} = 1 $
since the toric description of 
the corresponding \cym s 
and their mirror manifolds
are well-known \cite{vb,hkty1}. 
%The $U(1)$ charge of the gound state 
%in the $j^{-l}$ twisted sector is obtained to be;
Using the results of Intriligator and Vafa \cite{iv},
we can construct the $(c,c)$ and $(a,c)$ rings,
where $c$ ($a$) denotes chiral (anti-ciral).
Also we could have the left and right $U(1)$ charges of 
the ground state \  
$\keta h$ \ 
in the $h$-twisted sector 
of the $(a,c)$ ring.
In terms of spectral flow, \ $\keta h$ is mapped to the (c,c)
state $\ketc {h^{\prime}}$ with $h^{\prime} = hj^{-1}$.
Then the charges of the (a,c) ground state of $h$-twisted sector
$ \keta {h} $
%${\Big|0\Big\rangle^{(h)}_{(a,c)}}$
are obtained to be
\begin{equation}
\label{uac}
\begin{array}{cc}
\left(\begin{array}{c}
J_{0} \\
\bar{J_{0}}
\end{array} \right) &
% {\Big|0\Big\rangle^{(h)}_{(a,c)}}
\end{array}
\keta {h} 
= 
\begin{array}{cc}
\left(\begin{array}{c}
{ - \sum_{\thd>0}{(1-q_{i}-\thd)}}
+ \sum_{\thd=0} {(2q_{i}-1)}
 \\
{ \sum_{\thd>0}{(1-q_{i}-\thd)}}
\end{array} \right) &
\keta {h}.
\end{array}
\end{equation}

%In the following we will focus on the $(-1,1)$ states 
%in $(a,c)$ ring, 
%where states are labelled by left and right $U(1)$ charges.
%These states can be identified with $(1,1)$ forms on \cym s
%and hence correspond to divisors which can be described
%in terms of toric geometry.

%In the following
%we will associate an integral point inside $\Delta^*(w)$,
%i.e. an exceptional divisor, with a $(-1,1)$ state 
%which can be written in the form $\keta {j^{-l}}$.
%To explain our method, 
First we consider the \lg\ models 
with the Fermat type superpotentials
\begin{equation}
W_{Fermat} = 
X_{1}^{a_{1}} +
X_{2}^{a_{2}} +
\cdots +
X_{n}^{a_{n}}, 
\end{equation}
and their $(a,c)$ states.
We define the phase symmetries $\rho_{i}$ which act on $X_{i}$ as
\begin{equation}
\rho_{i}X_{i} = e^{-2 \pi i q_{i} }X_{i},
\end{equation}
with trivial action for other fields.
The operator $ \rho_{i} $
can be represented by a diagonal matrix
whose diagonal matrix elements are 1 except for
$ {(\rho_{i})}_{i,i} 
= e^{-2 \pi i q_{i} }$.
It is obvious that
\begin{equation}
j^{-1} = \rho_{1} \cdots \rho_{5}.
\end{equation}

In ref.\cite{sa2} the mirror map
for the $(a,c)$ ground states 
in the $j^{-l}$-twisted sector
$\keta {j^{-l}}$  
are considered.
In the $j^{-l}$-twisted sector, if a field $X_{i}$ is invariant 
then 
\begin{equation}
\rho_{i}^{-l} = \rho_{i}^{-l_{i}} = {\rm identity},
\end{equation}
where $-l_{i} \equiv -l \ {\rm mod} \ a_{i}$ and one gets
\begin{equation}
\label{key1}
j^{-l} = \prod_{{-l{q_{i}}} \notin \ZZ} {\rho_{i}^{l_{i}}}.
\end{equation}
So, we may represent
${\keta {j^{-l}}}
=\keta {\prod_{{-lq_{i}} \notin \ZZ}{\rho_{i}^{{l_{i}}}}}$.
Furthermore, 
we can calculate the $U(1)$ charges of 
this ground state 
and the result is 
\begin{equation}
{( - \sum_{{-lq_{i}} \notin \ZZ}{l_{i}q_{i}}, 
\sum_{{-lq_{i}} \notin \ZZ}{l_{i}q_{i}})}. 
\end{equation}
Using the phase symmmetry $\rho_{i}$ considered above,
one can obtain the following mirror map for \lg\ model
\cite{sa2};
\begin{equation}
\keta {\prod_{{-lq_{i}} \notin \ZZ}{\rho_{i}^{l_{i}}} } \ \ \
\stackrel{\rm mirror \ partner}{\longleftrightarrow} \ \ \
\prod_{{-lq_{i}} \notin \ZZ}{X_{i}^{l_{i}}}.
\end{equation}

There are potentials of other type,
i.e. so-called chain (tadpole) potential
\begin{equation}
W_{chain} = X_{1}^{a_{1}}X_{2} + \cdots +
X_{n-1}^{a_{n-1} - 1}X_{n} + X_{n}^{a_{n}}
\end{equation}
and loop potential
\begin{equation}
%W_{loop} = \sum_{i = 1}^{n}{X_{i}^{a_{i}}X_{i+1}}
W_{loop} = X_{1}^{a_{1}}X_{2} + \cdots + X_{n}^{a_{n}}X_{1}.
\end{equation}
In general the \lg\ superpotentials contain 
Fermat, chain and loop terms.
To study the mirror theory,
we have to consider the transposed potentials \cite{bh}.
For the chain potentials,
their transposed potentials are
\begin{equation}
\bar{W}_{chain} = \bar{X_{1}}^{a_{1}} + \bar{X_{1}}  
\bar{X_{2}}^{a_{2}} 
+ \cdots +
{\bar{X}}_{n-1} \bar{X_{n}}^{a_{n}},
\end{equation}
and for the loop potentials
\begin{equation}
%W_{loop} = \sum_{i = 1}^{n}{X_{i}^{a_{i}}X_{i+1}}
\bar{W}_{loop} = \bar{X_{1}} \bar{X_{2}}^{a_{1}} + \cdots 
+ {\bar{X}}_{n-1} \bar{X_{n}}^{a_{n-1}}+ \bar{X_{n}}  
\bar{X_{1}}^{a_{n}}.
\end{equation}
Their corresponding defining equations
are the hypersurfaces of the mirror manifold.
For the Fermat type potentials 
the transposed potentials
are the same form
as the original ones.

The phase symmetreis $\rho_{i}$ of chain and loop potentials 
were studied by Kreuzer \cite{kreuzer}.
Let us briefly summarize the results.
The definition of phase transformations is
\begin{equation}
\rho_{i}X_{l} = {\rm exp(2 \pi i}\vp_{l}^{(i)}{\rm )}X_{l}.
\end{equation}

For the chain potentials, we have
\begin{equation}
\vp_{l}^{(i)} = \frac{(-1)^{l-i+1}}{a_{i} \cdots a_{l}}
\ \ {\rm for} \ 1 \le l \le i,
\end{equation}
and $\vp_{i+1}^{(i)} = \cdots = \vp_{n}^{(i)} = 0$.
Moreover we have the following relations,
\begin{equation}
a_{l}\vp_{l}^{(i)} + \vp_{l+1}^{(i)} = - \delta_{l}^{i},
\end{equation}
\begin{equation}
a_{i+1}\vp_{l}^{(i+1)} + \vp_{l}^{(i)} = - \delta_{l}^{i+1}.
\end{equation}
From these relations, we can see
\begin{equation}
\label{chain}
(\rho_{i})^{a_{i}} \rho_{i-1} = 1,
\end{equation}
\begin{equation}
q_{i} = - \sum_{j=i}^{n}{\vp_{i}^{(j)}}.
\end{equation}

For the loop potentials, one obtains
\begin{equation}
\vp_{i+l}^{(i)} = \frac{(-1)^{n-l}a_{i} \cdots a_{i+l-1}}{\Gamma}
\ \ {\rm for} \ 0 \le l < n,
\end{equation}
where $\Gamma = A - (-1)^{n},\ \ A = a_{1}a_{2} \cdots a_{n}$.
In a similar fashion as chain potentials,
one can see
\begin{equation}
a_{l}\vp_{l}^{(i)} + \vp_{l+1}^{(i)} = - \delta_{l+1}^{i},
\end{equation}
\begin{equation}
a_{i}\vp_{l}^{(i+1)} + \vp_{l}^{(i)} = - \delta_{l}^{i}
\end{equation}
and
\begin{equation}
\rho_{i} (\rho_{i+1})^{a_{i}} = 1,
\end{equation}
\begin{equation}
q_{i} = - \sum_{l=1}^{n}{\vp_{i}^{(l)}}.
\end{equation}

For both cases we may represent any twist $h$
uniquely in the form
\begin{equation}
h =
\prod
{\rho_{i}^{\al_{i}}}
\end{equation}
where $0 \le \al_{i} <  a_{i}$.
A simple calculation shows that 
the $U(1)$ charge of the state $\keta h$ is
\begin{equation}
(-\sum_{{\rm all} \ i}{\varphi_{i}},
\sum_{{\rm all} \ i}{\varphi_{i}})
\end{equation}
where 
${\varphi_{i}} = 
\sum_{l=i}^{n}{{\al_{l}}{\varphi_{i}^{(l)}}}$.

%Using the phase symmmetry $\rho_{i}$ considered above,
Due to the obsevation of the $U(1)$ charge,
one can obtain the following mirror map for \lg\ model
\cite{kreuzer};
\begin{equation}
\label{lgmirror}
\keta {\prod
%_{{-lq_{i}} \notin \ZZ}
{\rho_{i}^{l_{i}}} } \ \ \
\stackrel{\rm mirror \ partner}{\longleftrightarrow} \ \ \
\prod
%_{{-lq_{i}} \notin \ZZ}
{\bar{X}_{i}^{l_{i}}},
\end{equation}
where $\bar{X_{i}}$ are the fields of the transposed potential.
The special class of this mirror symmetry, namely
\begin{equation}
\label{specialmirror}
{\rm  (-1,1) \ states} \ \ \keta {j^{-l}} \ \ \
\stackrel{\rm mirror \ partner}{\longleftrightarrow} \ \ \
{\rm  (1,1) \ states} \ \ 
\prod_{{\varphi_{i}} \notin \ZZ}{\bar{X}_{i}^{l_{i}}} \keta {0},
\end{equation}
is important since the monomials $\prod_{i}{\bar{X}_{i}^{l_{i}}}$
correspond to the monomial deformations
of the complex structure of the mirror manifold.

\subsection{Correspondence between \lg\ model and toric geometry}

In this section,we study the toric geometrical structure
of  \lg\  model.
We will show that the Batyrev's formula of
eqs.(\ref{vb11}) and (\ref{vb21})
exactly count the number of 
the $(-1,1)$ and $(1,1)$ states, respectively,
as expected.
This analysis enables us to tell 
that which states sould correspond to
the twisted sector problem.

First, we consider the Fermat type potentials
and study their toric aspects.
This was already done in \cite{sa3},
so we briefly review the discussion of \cite{sa3}
and make it more precise.
Then the generalization to the chain and loop potentials
will be given.

The eq.(\ref{key1}) is the key equation for our purpose. 
This implies
\begin{equation}
\label{key2}
-lq_i = u_i - l_{i}q_{i} \quad \quad {\rm for} \ 
i=1 \sim 5,
\end{equation}
where $-l_i$ are defined to be
$-a_i+1 \le -l_{i} \le 0$.
Thus $u_i$ are uniquely determined.
Clearly, $u_i$ are integers
and $l_{i}=0$ if $X_{i}$ is invariant
under $j^{-l}$ action.
%(see eq.(\ref{key1})).
$u_1$ always vanish
since $w_1 = 1$.

Let $u \equiv (u_2,u_3,u_4,u_5)$
be an integral vector.
It can be shown \cite{sa3}
that $u$ is on a face 
of a dual polyhedron $\Delta^*(w)$.
So we should assert
that $u$ is just an integral point inside
$\Delta^*(w)$,
which can be identified with the exceptional divisors.
Through this identification,
we obtain the one-to-one correspondence
between the $(-1,1)$ state $\keta {j^{-l}}$
and the exceptional divisor.
Moreover, once this identification is made,
we can obtain the monomial-divisor mirror map
\cite{agm3}
for \cym s.
%For more details, see
%refs.\cite{sa2,sa3,sa4}.

A simple observation shows that
the above identification is equivalent to
the following identification
\begin{equation}
%\label{rhotoric1}
\rho_{i}^{a_{i}} = 1  \  \  \rightarrow
\ \ (0, \cdots,1, \cdots,0)
\in \Delta^{*},
\end{equation}
where the components of the vectors are
zero except for the $i-$th entry being 1.
More concretely, for the twist $j^{-l}$ 
eq.(\ref{key2}) implies
\begin{equation}
\label{rhotoric}
\rho_{i}^{l} = 
{(\rho_{i}^{a_{i}})}^{-u_{i}}{\rho_{i}^{l_{i}}}.
\end{equation}

If the $(-1,1)$ state
$\keta {j^{-l}}$ with 2 and 3 invariant field under 
$j^{-l}$ action,
the toric data $u$ obtained by the above method
lies on the 2 and 1 dimensional face $\Theta^{*}$,
respectively.
If the $(-1,1)$ state $\keta {j^{-l}}$
has only one invariant field,
we can obtain the toric data 
$\nu^{*(\alpha)}$
by the same mehtod
and $\nu^{*(\alpha)}$ lies on a codimension 1
(i.e. 3-dimensional) face $\Theta^{*}$.
Moreover, in this case we can further obtain 
the new toric data
$\nu^{*(\beta)}$
from the descendent twist $j^{-l+1}$.
If only one field is invariant under $j^{-l+1}$ action,
we can obtain the third toric data 
$\nu^{*(\gamma)}$
from the twist $j^{-l+2}$.
This successive process will stop 
when the twist $j^{-l+n}$ has 2 or 3 invariant fields.
These results are 
in complete agreement with
the subtraction of 
$\sum_{\bf{codimension} \; {\Theta^{*}} = 1}
{l^{'}({\Theta^{*}})}$
in the Batyrev's formula
(\ref{vb11}).

It was noticed in \cite{sa3} 
that the mirror partners of the $(-1,1)$ states
written in the form
$\prod_{ \thd= 0}{{X_{i}}^{l_{i}}}
\keta{h}$
can not be described by the monomials of fields
in the transposed potential.
So it is natural to expect 
that  these states correspond to the twisted sector problem.
To prove this, we should note the fact that
the number of $(-1,1)$ states
%It is shown in \cite{sa3} that the number of $(-1,1)$ states
$\keta {j^{-l}}$ with two invariant fields
(say, $X_{i}$ and $X_{j}$) is $l^{'}({\Theta}^{*})$,
where ${\Theta}^{*}$ is the 2-dimensional dual face of
1-dimensional face ${\Theta}(\nu_{i}, \nu_{j})$.
The toric data 
$\nu_{i}$ is the vector whose components are $-1$ 
except for the $i-$th entry being $a_{i}-1$.
%The toric data $\nu_{i}$ 
This vector
corresponds to the monomial ${X_{i}}^{a_{i}}$.
If ${\Theta}(\nu_{i}, \nu_{j})$ contains integral points
in its interior,
these points should correspond to the monomials
${X_{i}}^{{\alpha}_{i}}X_{j}^{{\alpha}_{j}}$,
and the number of such monomials is shown to be 
$l^{'}(\Theta)$.
Since the $U(1)$ charge of the state 
$\keta {j^{-l+1}}$
with two invariant fields is
$(-2+\sum_{lq_{i}\in \ZZ }{q_{i}},\sum_{lq_{i}\in \ZZ }{q_{i}})$
and 
${{\alpha}_{i}q_{i}}+{{\alpha}_{j}q_{j}} = 1$,
the states 
${X_{i}}^{{\alpha}_{i}-1}{X_{j}}^{{\alpha}_{j}-1}
\keta {j^{-l+1}}$
are $(-1,1)$ states.
One can see that the number of such states is 
$l^{'}({\Theta})l^{'}({\Theta}^{*})$
with 2-dimensional dual face
${\Theta}^{*}$.

To generalize the above result 
to chain and loop potentials,
%we should adopt the following identification,
we should extend the identification of 
(\ref{rhotoric}) to
\begin{equation}
\label{rhotoric1}
\rho_{i}^{a_{i}} = 1  \  \  \rightarrow
\ \ (0, \cdots,1, \cdots,0)
\in \Delta^{*}
\end{equation}
and
\begin{equation}
\label{rhotoric2}
\rho_{i}^{a_{i}}{\rho_{j}} = 1  \  \  \rightarrow
\ \ (0, \cdots,1, \cdots,0)
\in \Delta^{*},
\end{equation}
%because of the \'{e}tale map of eq.(\ref{etale}),
where the components of the vectors are
zero except for the $i-$th entry being 1
as before.
It can be shown that 
these identifications coincide with the equivalence
between the mirror map of (\ref{lgmirror})
and the \'{e}tale map of eq.(\ref{etale}).

Furthermore we can point out 
the relation between the Mori vectors
and the phase symmetries.
A Mori cone which consists of Mori vectors 
is a dual cone of a K\"ahler cone 
and can describe the good coordinates 
for a moduli space of complex structure \cite{hkty1}.
A Mori vector represents a relation of the monomials
in a defining equation of a mirror \cym .
Since the mirror map (\ref{specialmirror}) 
connects these monomials
and the $j^{-l}$ 
twist,
we can expect that a Mori vector translates into
a relation of $j^{-l}$ 
or ${\rho_{i}}^{l_{i}}$ twists.
Indeed, a following type of  relation,
\begin{equation}
\prod_{i}{\rho_{i}^{l_{i}}} = {\rm identity},
\end{equation}
leads to a Mori vector.
In the later section we will show some examples.

\section{Mirror symmetry and the web of string vacua}

\subsection{Twisted sector problem revisited}

It is a well known fact \cite{vb,hkty1} that in certain situations 
not all of the K\"ahler moduli space can be described 
by toric divisors of the \cym\ ${\cal M}$.
Similarly, the deformation of the defining equation
of the hypersurface of ${\cal M}$ in a toric variety is 
in general not sufficient to describe 
the complex structure moduli space of ${\cal M}$.
More precisely,
as remarked under (\ref{vb11}) and (\ref{vb21}),
there are non toric divisors in ${\cal M}$ when
\begin{equation}
\label{tw11}
\sum_{\bf{codimension} \; \Theta^{*} = 2, 
\Theta^{*} \in \Delta^{*}}
{l^{'}(\Theta^{*})
l^{'}(\Theta)}
\end{equation}
is non-zero.
In a similar fashion, when
\begin{equation}
\label{tw21}
\sum_{\bf{codimension} \; \Theta = 2, 
\Theta \in \Delta}
{l^{'}(\Theta)
l^{'}(\Theta^{*})}
\end{equation}
is non-zero 
there are non-algebraic deformations 
of the complex structure deformations 
of the complex structure of ${\cal M}$.
%These deformations cannot be treated by toric geometry,
%so that we can study only a part of moduli space of $X$
%and $X^{*}$.

In section 3.2, we have shown that in the \lg\ context 
these numbers (\ref{tw11}) ,(\ref{tw21}) 
are equivalent to the number of 
%the $(-1,1)$ sates, 
%$\prod_{\thd} \in \ZZ}{X_{i}^{l_{i}}}$,
%the $(-2,1)$ states 
the states ${\prod_{{\thd} \in \ZZ}{X_{i}^{l_{i}}}}\keta {j^{-l}}$ 
with $U(1)$ charge $(-1,1)$, $(1,1)$
respectively.

\subsection{Orbifold construction and the web of \lg\ vacua}

In this section, we would like to propose a method 
which circumvent 
this twisted sector problem 
under some circumstances.
As a result 
either all of the non-toric divisors in (\ref{tw11})
or the non-algebraic deformations 
of the complex structure in (\ref{tw21}),
can be treated by this method.
In the following we will concentrate ourselves on
the K\"{a}hler moduli space
and the corresponding $(-1,1)$ states.

The main idea is orbifold construction,
i.e. we further divide the original theory 
by a new discrete symmetry,
whose generator is denoted by $g$, 
and project out the unexpected states
${\prod_{{\thd} \in \ZZ}{X_{i}^{l_{i}}}}\keta {j^{-l}}$
with $U(1)$ charge $(-1,1)$.
At the same time,
we have new twisted states
$\keta {g^{-m}j^{-l^{'}}}$
with charge $(-1,1)$,
which can be identified with toric data.
According to this orbifoldization, 
the superpotential should be deformed,
as discussed in \cite{ls}

On the toric geometry side, after this procedure is done
we have new polyhedra ${\tilde{\Delta}}^{*}$ and ${\tilde{\Delta}}$
such that
\begin{equation}
{\tilde{\Delta}}^{*}
\supset
{{\Delta}^{*}},
\ \ 
{\tilde{\Delta}}
\subset
{\Delta}.
\end{equation}
This relation implies that 
the moduli spaces of the original theory
and the orbifoldized thory are connected
\cite{ca1,ca}.

Can we find such a suitable twist?
In the following, we  will show that
under some circumstances
we can find the appropriate $g$ for our construction.

Consider the state
$\prod_{ \thd= 0}{{X_{i}}^{\alpha_{i}}}
\keta{j^{-l+1}}$
where $0 \le \alpha_{i} \le {{a_{i}}-2}$.
The $U(1)$  charge of this state
is obtained to be
\begin{equation}
\label{uac1}
%\begin{array}{cc}
\left(\begin{array}{c}
{ - \sum_{\thd>0}{(1-q_{i}-\thd)}}
+ \sum_{\thd=0} {(2q_{i}-1)}
+\sum_{\thd=0} {{\alpha_{i}}{q_{i}}}
 \\
{ \sum_{\thd>0}{(1-q_{i}-\thd)}
+\sum_{\thd=0} {{\alpha_{i}}{q_{i}}}
}
\end{array} \right) ,
%\end{array}
\end{equation}
where $h = j^{-l}$ and $h^{'} = h{j^{-1}}$.

%\begin{equation}
%\end{equation}

In this case the appropriate $g$
is found to be
\begin{equation}
\label{defg}
g^{-1} = 
\prod_{\thd = 0}{{\rho_{i}}^{{{\alpha}_{i}}+1}}.
\end{equation}
Using this $g$,
the new twisted state 
$\keta{{g^{-1}}{j^{-l+1}}}$
will appear and its $U(1)$ charge is obtained to be
$(-Q,Q)$, 
where
$Q = 
{  \sum_{\thd>0}{(1-q_{i}-\thd)}}
+ \sum_{\thd=0} {{\alpha_{i}}{q_{i}}}.
$
%\begin{equation}
%\end{equation}

We see that if
\begin{equation}
\sum_{\thd = 0}{(2q_{i} - 1)} +
\sum_{\thd = 0}{{\alpha_{i}}{q_{i}}}
=
- \sum_{\thd = 0}{{\alpha_{i}}{q_{i}}},
\end{equation}
then the two states
$\prod_{ \thd= 0}{{X_{i}}^{\alpha_{i}}}
\keta{j^{-l+1}}$
and
$\keta{{g^{-1}}{j^{-l+1}}}$
have same $U(1)$ charge.
To obtain the toric data of the orbifoldized theory,
we should use the identifications of (\ref{rhotoric1})
and (\ref{rhotoric2}) in $j^{-l+1}$.

%\subsection{Verifications}
\subsection{Examples}

As a first example, we consider the \lg\ model with the  
superpotential
\begin{equation}
\label{w21}
W_{1} = 
X_{1}^{12}+X_{2}^{12}+X_{3}^{12}+X_{4}^{4}+X_{5}^{2},
\end{equation}
with $U(1)$ charges of $X_{i}$ being
\begin{equation}
( \
\frac{1}{12}, \
\frac{1}{12}, \
\frac{1}{12}, \ 
\frac{1}{4}, \
\frac{1}{2} \
),
\end{equation}
which corresponds to the hypersurface
in $WCP_{(1,1,1,3,6)}^{4}[12]$ and were studied in ref.\cite{klemm}.

We find that there are three $(-1,1)$ states
$\keta {j^{-2}}$,
$\keta {j^{-4}}$ and 
$X_{4} {\keta {j^{-3}}}$.
From the first two of these states
we can easily obtain the toric data and
the results are displayed in Table \ref{w1t},
where Dim. means the dimension of the dual face 
on which the point $\nu^{*}$ lies.
The reader should remind 
that for \cy\ threefolds,
a point lying on a face of Dim. = 1 or 2
corresponds to a toric divisors
and 
$\nu_0^*=(0,0,0,0)$
corresponds to the canonical divisor.
Note that since only one field is invariant under 
the twist $j^{-2}$,
we should consider the descendent twist $j^{-1}$
to obtain the toric data completely,
as explained.

\begin{table}[htbp]
\[ \begin{tabular}{||c|c|c|c||} \hline
(a,c) state &
twist & $\nu_j^*$ vector & Dim. \\ \hline \hline
$\keta {j^{-2}}$ & 
${j^{-2}}$ &
$\nu_6^*=(0,0,0,-1)$ & 
$ 3 $ 
\\ \hline
 & 
${j^{-1}}$ &
$\nu_0^*=(0,0,0,0)$ & 
$ 4 $ 
\\ \hline
$\keta {j^{-4}}$ & 
${j^{-4}}$ &
$\nu_7^*=(0,0,-1,-2)$ & 
$ 2 $ 
\\ \hline
\end{tabular} \]
\caption{The toric data of
the $(-1,1)$ states
in the 
\lg\ orbifold of $W_{1}$}
\label{w1t}
\end{table}

The remaining $(-1,1)$ state represented by
$X_{4} {\keta {j^{-3}}}$
should correspond to so-called 
twisted sector problem in toric geometry.
%, i.e. the corresponding divisor cannot be identifid with 
%a point in $\Delta^{*}(w)$.
%Only the number of such non-toric divisors
%can be described by toric geometry \cite{vb}.
%We can show \cite{sa4} that for Fermat type models the number of 
%these $(-1,1)$ states is equal to the number of such non-toric  
divisors.
%Using these understandings, we will propose an 
%effective 
%useful method
To solve this twisted sector problem,
we will use the mehtod proposed in the section 4.2.
Let us remind that
the essential ingredient is an orbifold construction.
If one finds an appropriate orbifoldization,
the $(-1,1)$ states such as 
$X_{4} {\keta {j^{-3}}}$
are projected out and new $(-1,1)$ states will appear
in the new twisted sectors.
For our example, we can find such an appropriate orbifoldization.

Consider the following superpotential
\begin{equation}
\label{w2}
W_{2} = 
Y_{1}^{12}+Y_{2}^{12}+Y_{3}^{12}+Y_{4}^{2}Y_{5}+Y_{5}^{2}.
\end{equation}
Note that the set of fields $Y_{i}$ have 
the same $U(1)$ charges as $X_{i}$'s.
So $W_{2}$-theory has the same topological numbers as  
$W_{1}$-theory.
Hence one may identify the fields $X_{i}$ and $Y_{i}$;
\begin{equation}
X_{i} = Y_{i} \ \ {\rm for} \ \ i = 1,2,3, \ \ \ 
X_{4}^{4} = Y_{5}^{2}, \ \ X_{5}^{2} = Y_{4}^{2}Y_{5}.
\end{equation}
For one-to-one identification, we should further identify
\begin{equation}
\label{z2}
X_{4} \sim -X_{4},\ \ \  
X_{5} \sim -X_{5}. 
\end{equation}
Since this twist is not included 
in the $U(1)$ twist $j^{-l}$,
 $W_{2}$-theoy can be considered as
$\ZZ_{2}$ orbifold of $W_{1}$-theory
\cite{ls}.
We denote by $g$ this $\ZZ_{2}$ twist.
From the results of section 4.2,
we find the appropriate $g$ of (\ref{defg})
is,
\begin{equation}
g^{-1} = {{\rho_{4}}^2}{\rho_{5}},
\end{equation}
this agrees with $\ZZ_{2}$ identification (\ref{z2}).
The new $(-1,1)$ state is 
$\keta {g^{-1}j^{-3}}$.
Below one finds our results.
%There is no 
%twisted sector problem as expected.
The twisted sector problem has resolved 
as expected.

\begin{table}[htbp]
\[ \begin{tabular}{||c|c|c|c||} \hline
(a,c) state & 
twists &
$\nu_j^*$ vector & Dim. \\ \hline \hline
$\keta {j^{-2}}$ & 
${j^{-2}}$ &
$\nu_6^*=(0,0,0,-1)$ & 
$ 3 $ 
\\ \hline
 & 
${j^{-1}}$ &
$\nu_0^*=(0,0,0,0)$ & 
$ 4 $ 
\\ \hline
$\keta {j^{-4}}$ & 
${j^{-4}}$ &
$\nu_7^*=(0,0,-1,-2)$ & 
$ 2 $ 
\\ \hline
$\keta {g^{-1}j^{-3}}$ & 
${g^{-1}j^{-3}}$ &
$\nu_{11}^*=(0,0,-1,-1)$ & 
$ 3 $ 
\\ \hline
  &
${g^{-1}j^{-2}}$ & 
$\nu_{10}^*=(0,0,-1,0)$ & 
$ 3 $ 
\\ \hline
  &
${g^{-1}j^{-1}}$ & 
$\nu_9^*=(0,0,-1,1)$ & 
$ 3 $ 
\\ \hline
  &
${g^{-1}}$ & 
$\nu_8^*=(0,0,0,1)$ & 
$ 1 $ 
\\ \hline
\end{tabular} \]
\caption{The toric data of
the $(-1,1)$ states in 
the \lg\ orbifold of $W_{2}$}
\label{w2t}
\end{table}

Our results shows that the toric data $(\Delta_{i}, \Delta_{i}^{*})$  
which 
correspond to $W_{i}$-theory satisfy
\begin{equation}
\Delta_{1} \supset \Delta_{2}, \ \ 
{\rm and} \ \ \Delta_{1}^{*} \subset \Delta_{2}^{*}.
\end{equation}
These properties of toric data imply that 
the mouli spaces of these two theories are connected 
\cite{ca1,ca}.
From our construction, the singular $\ZZ_{2}$ orbifold point
should be considered that
the superpotential is $W_{2}$ and $(-1,1)$ states are
$\keta {j^{-2}}$ and
$\keta {j^{-4}}$.
Thus on the singular orbifold point,
the corresponding toric data are 
$(\Delta_{2}, \Delta_{1}^{*})$.
This is not a dual pair
so that one could not obtain a smooth \cym .
However, this is a boundary point which connects 
the moduli spaces of the two theories described by
the superpotentials $W_{1}$ and $W_{2}$.
Our orbifold construction corresponds to the toric construction
in refs.\cite{klemm,ca},
in which the structures of moduli spaces of \cym s 
are studied.
%Details and some generalizations will appear elsewhere \cite{sa4}.

Then, we consider the relation 
between Mori vectors and the twist operators.
In the original theory whose potential is $W_{1}$,
there is 
a $(-1,1)$ state $\keta {j^{-4}}$ 
whose toric data is $\nu_{7}^{*}$.
In terms of the mirror map for \lg\ model
(\ref{specialmirror}), the relation
$({j^{-4}})^{3} = {j^{-12}} = {\rm identity}$
translates into
$({X_{1}}^{4}{X_{2}}^{4}{X_{3}}^{4})^{3}
 = ({X_{1}}^{12})({X_{2}}^{12})({X_{3}}^{12})$.
This implies the Mori vector 
$l_{1}^{(1)} = 
(0;1,1,1,0,0,-3,0).$
In a similar manner, 
${j^{-4}} = (j^{-1})^{4}$
translates into
$({X_{1}^{4}}{X_{2}^{4}}{X_{3}^{4}})(X_{4}^{4})
(X_{5}^{2})^{2}
= {({X_{1}}{X_{2}}{X_{3}}{X_{4}}
{X_{5}})}^{4}$
and we get the Mori vector
$l_{1}^{(2)} = 
(-4;0,0,0,1,2,1,0).$

In the orbifoldized theory
whose potential is $W_{2}$,
we have a new operator set 
$\{ {\tilde{\rho}}_{i} \}$.
From the eq.(\ref{chain}), we have
\begin{equation}
\label{rhoalg2}
{\tilde{\rho_{1}}}^{12} = {\tilde{\rho_{2}}}^{12}
= {\tilde{\rho_{3}}}^{12} = {\tilde{\rho_{4}}}^{2}
= {\tilde{\rho_{4}}}{\tilde{\rho_{5}}}^{2}
= {\rm identity}. 
\end{equation}
Note that from these basic relations of 
the operators ${\tilde{\rho_{i}}}$,
we have the new relation 
%${\tilde{\rho_{5}}}^{4} = {\rm identity}$. 
$ {({{\tilde{\rho}}_{4}}{{\tilde{\rho}}_{5}}^{2})}^{2}
{({\tilde{\rho_{4}}}^{2})}^{-1} = 
{{\tilde{\rho}}_{5}}^{4} = 
{\rm identity}$.
This relation 
$ {({{\tilde{\rho}}_{4}}{{\tilde{\rho}}_{5}}^{2})}^{2}
{({\tilde{\rho_{4}}}^{2})}^{-1} = 
{{\tilde{\rho}}_{5}}^{4} = 
{\rm identity}$
gives us the new Mori vector.
%where ${{\tilde{\rho}}_{i}}$ are 
%the phase symmetry of $W_{2}$.
In terms of the mirror map for \lg\ model
(\ref{specialmirror}), 
we can obtain
$({\bar{Y_{4}}}{\bar{Y_{5}}}^{2})^{2}
= ({\bar{Y_{4}}}^{2})
({\bar{Y_{5}}}^{4}).
$
The corresponding Mori vector is 
$l_{2}^{(1)} = 
(0;0,0,0,1,-2,0,1).$
In this orbifold model, 
${j^{-4}} = (j^{-1})^{4}$
translates into the Mori vector
$l_{2}^{(2)} = 
(-4;0,0,0,0,4,1,-1).$
As a result, the Mori vectors satisfy the relation
$l_{1}^{(2)} = l_{2}^{(1)} + l_{2}^{(2)}$
\cite{klemm}.
We can understand this relation 
as the change of the set of twist operators 
$\{ {\rho}_{i} \}$
of the phase symmetry,
or equivalently, as the orbifoldization.

As a second example, we consider 
the hypersurface in $WCP_{(1,2,3,6,6)}[18]$
which was considered in \cite{agm,sa3}.
In this case the orbifoldized potential 
contains a loop part.
The defining superpotential is
\begin{equation}
\label{w3}
W_{3} = 
X_{1}^{18}+X_{2}^{9}+X_{3}^{6}+X_{4}^{3}+X_{5}^{3}.
\end{equation}
There are seven $(-1,1)$ states
and two of them correspond to
the twisted sector problem, i.e. 
$X_{4} \keta{j^{-2}}$
and
$X_{5} \keta{j^{-2}}$.
The appropriate twist is
\begin{equation}
g^{-1} = {\rho_{4}}^{2}{\rho_{5}},
\end{equation}
and this generates the $\ZZ_{3}$ being
\begin{equation}
X_{4} \sim \alpha^{2} X_{4},\ \ \  
X_{5} \sim \alpha X_{5} 
\end{equation}
where $\alpha$ is a cubic loot of unity.
After the orbifoldization, the deformed superpotential includes 
a loop potential and obtained to be
\begin{equation}
\label{w4}
W_{4} = 
Y_{1}^{18}+Y_{2}^{9}+Y_{3}^{6}+Y_{4}^{2}Y_{5}+Y_{4}Y_{5}^{2}.
\end{equation}

In the orbifoldized theory,
the new $(-1,1)$ states are
$ \keta {g^{-1}j^{-2}}$ and
$ \keta {g^{-2}j^{-2}}$. 
The corresponding toric data are
displayed in Table \ref{w4t}
with the descendent twists.
Our results show that
the twisted sector problem 
has circumvented as expected.

\begin{table}[htbp]
\[ \begin{tabular}{||c|c|c|c||} \hline
(a,c) state & 
twists &
$\nu_j^*$ vector & Dim. \\ \hline \hline
$ \keta {g^{-1}j^{-2}}$ & 
${g^{-1}j^{-2}}$ & 
$\nu_6^*=(0,0,-1,0)$ & 
$ 3 $ 
\\ \hline
  &
${g^{-1}j^{-1}}$ & 
$\nu_8^*=(0,0,1,-1)$ & 
$ 3 $ 
\\ \hline
  &
${g^{-1}}$ & 
$\nu_{10}^*=(0,0,1,0)$ & 
$ 1 $ 
\\ \hline
$\keta {g^{-2}j^{-2}}$ & 
${g^{-2}j^{-2}}$ & 
$\nu_7^*=(0,0,0,-1)$ & 
$ 3 $ 
\\ \hline
  &
${g^{-2}j^{-1}}$ & 
$\nu_9^*=(0,0,-1,1)$ & 
$ 3 $ 
\\ \hline
  &
${g^{-1}}$ & 
$\nu_{11}^*=(0,0,0,1)$ & 
$ 1 $ 
\\ \hline
\end{tabular} \]
\caption{The toric data of
\lg\ orbifold of $W_{4}$}
\label{w4t}
\end{table}

Our third example \cite{ca1} is 
the hypersurface in $WCP_{(1,2,2,2,7)}[14]$.
The corresponding superpotential is
\begin{equation}
\label{w5}
W_{5} = 
X_{1}^{14}+X_{2}^{7}+X_{3}^{7}+X_{4}^{7}+X_{5}^{2},
\end{equation}
and there are two $(-1,1)$ states
$\keta {j^{-2}}$ and $\keta {j^{-8}}$.
%There is one $(-1,1)$ state which corresponds to 
%the twisted sector problem.
Once we identify
\begin{equation}
\label{identifyw5}
%X_{2} \sim -X_{2} \ \ 
%X_{3} \sim -X_{3} \ \
X_{1} \sim -X_{1}, \ \ 
X_{5} \sim -X_{5},
\end{equation}
then we obtain the orbifoldized potential
\begin{equation}
\label{w6}
W_{6} = 
Y_{1}^{7}+Y_{2}^{7}+Y_{3}^{7}+Y_{4}^{7}+Y_{1}Y_{5}^{2}.
\end{equation}
The corresponding hypersurface is 
in $WCP_{(1,1,1,1,3)}[7]$.
This theory includes
two $(-1,1)$ states
$\keta {j^{-1}}$ and $\keta {j^{-2}}$.
%Since the identification (\ref{identifyw5})
%is a part of projective equivalnce of $WCP_{(1,2,2,2,7)}[14]$,
%\begin{equation}
%(X_{1},X_{2},X_{3},X_{4},X_{5})
% \sim ({\lambda}X_{1},{\lambda}^{2}X_{2},
%{\lambda}^{2}X_{3},{\lambda}^{2}X_{4},{\lambda}^{7}X_{5}),
%\end{equation}
%with ${\lambda} = -1$, 
%these two theories should be equivalent.
Though their weights are diferent,
these two theories should be equivalent
since the identification (\ref{identifyw5})
is a part of projective equivalnce of $WCP_{(1,2,2,2,7)}[14]$,
\begin{equation}
(X_{1},X_{2},X_{3},X_{4},X_{5})
 \sim ({\lambda}X_{1},{\lambda}^{2}X_{2},
{\lambda}^{2}X_{3},{\lambda}^{2}X_{4},{\lambda}^{7}X_{5}),
\end{equation}
with ${\lambda} = -1$.

The vertices of the dual polyhedron $\Delta_{5}^{*}$
corresponding to the $W_{5}$-theory are
\begin{eqnarray*}
&(-2,-2,-2,-7)&,  \ \
(1,0,0,0), \ \
(0,1,0,0), \\
&(0,0,1,0)&, \ \
(0,0,0,1), \ \
(0,0,0,-1), \\
&(-1,-1,-1,-3)&, \ \
(-1,-1,-1,-4).
\end{eqnarray*}
The toric data $(0,0,0,-1)$ and
$(-1,-1,-1,-4)$ correspond to
the two $(-1,1)$ states
$\keta {j^{-2}}$ and $\keta {j^{-8}}$, respectively.
The toric data $(-1,-1,-1,-3)$ can be identified
with the descendent twist $j^{-7}$.

Turning our attention to the $W_{6}$-theory,
the vertices of $\Delta_{6}^{*}$ are
\begin{eqnarray*}
&(-1,-1,-1,-3)&,  \ \
(1,0,0,0), \ \
(0,1,0,0), \\
&(0,0,1,0)&, \ \
(0,0,0,1).
\end{eqnarray*}

%At first sight, the (anti-) chiral structures of 
At first sight, the sets of toric data for
these two theories are different.
However, we can show that 
the toric data 
%which are obtained in terms of our method
which are obtained from $W_{6}$-theory
are the same as 
the ones from $W_{5}$-thory.
The results are displayed in Table \ref{w6t}

\begin{table}[htbp]
\[ \begin{tabular}{||c|c|c|c||} \hline
(a,c) state & 
twists &
$\nu_j^*$ vector & Dim. \\ \hline \hline
$ \keta {j^{-2}}$ & 
${j^{-2}}$ & 
$\nu_6^*=(0,0,0,-1)$ & 
$ 3 $ 
\\ \hline
  &
${j^{-1}}$ & 
$\nu_7^*=(-1,-1,-1,-4)$ & 
$ 3 $ 
\\ \hline
  &
${j^{0}}$ & 
$\nu_8^*=(-2,-2,-2,-7)$ & 
$ 1 $ 
\\ \hline
\end{tabular} \]
\caption{The toric data of
the $(-1,1)$ states in
the \lg\ orbifold of $W_{6}$}
\label{w6t}
\end{table}

As a final example,
we consider the hypersurface 
in $WCP_{(1,1,1,2,5)}[10]_{-288}^{1}$,
where the superscript denotes the number of $(1,1)$ forms
and the subscript implies the Euler number.
The corresponding superpotential is 
\begin{equation}
\label{w7}
W_{7} = 
X_{1}^{10}+X_{2}^{10}+X_{3}^{10}+X_{4}^{5}+X_{5}^{2}.
\end{equation}
This theory contains only one $(-1,1)$ state
$\keta {j^{-1}}$ 
whose toric data is the origin 
$(0,0,0,0)$
in the dual polyhedron ${\Delta}^{*}$.
The identification
\begin{equation}
\label{identifyw7}
X_{3} \sim -X_{3}, \ \ 
X_{5} \sim -X_{5},
\end{equation}
leads to the orbifoldized theory whose potential is
\begin{equation}
\label{w8}
W_{8} = 
Y_{1}^{10}+Y_{2}^{10}+Y_{3}^{5}+Y_{4}^{5}+Y_{3}Y_{5}^{2},
\end{equation}
which corresponds to the hypersurface
in $WCP_{(1,1,2,2,4)}[10]_{-192}^{3}$.
This theory contains the additional
two $(-1,1)$ states, i.e.
$\keta {j^{-2}}$
and
$\keta {j^{-5}}$
whose toric data are
$(0,0,0,-1)$ and
$(0,-1,-1,-2)$, respectively.
Our results implies that 
%$(\Delta_{i}, \Delta_{i}^{*})$ which 
%correspond to $W_{i}$-theory satisfy
\begin{equation}
\Delta_{{7}} \supset \Delta_{{8}}, \ \ 
{\rm and} \ \ \Delta_{{7}}^{*} \subset \Delta_{{8}}^{*}.
\end{equation}
%These properties of toric data imply that 
These relations show that
the moduli spaces of these two theories are connected,
even though their numbers of states,
namely their mass spectra,  
are different.
This should be considered as 
an example of the extremal transition
between the topologically different
\cym s.

\section{Application to type II strings}

In the context of string duality,
the $K3$-fibred \cym s are actvily studied.
The conjectured duality \cite{kv} is 
that the heterotic string on $K3 \times {T}^{2}$
is dual to the type II string on the \cym\ 
which admits $K3$-fibration.
Such type II strings are the promising candidates 
to obtain the Seiberg-Witten gauge theory \cite{seiwit}
by taking the suitable low energy limit \cite{kkv,w,kmv}.

In this section, we consider the simplest $K3$-fibred \cym s
constructed in \cite{klm},
i.e. the Fermat type \cym s in 
$WCP_{(1,1,2w_{3},2w_{4},2w_{5})}[2d]$
whose $K3$-fibre are
$WCP_{(1,w_{3},w_{4},w_{5})}[d]$,
where $d = 1 + w_{3} + w_{4} + w_{5}$.
For these models, 
%we can always obtain the $(-1,1)$ state
our \lg\ analysis shows that 
%there always exist 
the $(-1,1)$ state
$\keta {j^{-d}}$ 
always exists
and that
this state corresponds to 
the toric data $\nu_{6}^* = (0,-w_{3},-w_{4},-w_{5})$.
Since the dual polyhedron ${\Delta}^{*}$ contains the two vertices
$\nu_{1}^{*} = (-1,-2w_{3},-2w_{4},-2w_{5})$
and
$\nu_{2}^{*} = (1,0,0,0)$,
we see $\nu_{6}^{*} = {\frac{1}{2}} (\nu_{1}^{*} + \nu_{2}^{*})$.
This corresponds to the relation
$(j^{-d})^{2} = j^{-2d} 
%= {\rm identity}
$.
In other words, 
$\nu_{1}^{*}$,
$\nu_{2}^{*}$ and
$\nu_{6}^{*}$ 
correspond to the monomials
$X_{1}^{2d}$,
$X_{2}^{2d}$ and
$X_{1}^{d}X_{2}^{d}$, respectively,
of the defining equation 
of the mirror manifold
on which a type IIB string 
gets compactified.
The existence of these monomials is  
necessary to obtain 
the Seiberg-Witten geometry \cite{kkv,w,kmv}.
%\begin{equation}
%\label{sw1}
%a_{1}z + a_{2}/z + a_{3} + \cdots .
%\end{equation}
More precisely,
the defining equation of the 
$K3$-fibred
Calabi-Yau manifold
can be written as
\begin{equation}
p = {\frac{1}{2d}}{X_{1}^{2d}}
+ {\frac{1}{2d}}{X_{2}^{2d}}
+ {\frac{1}{d \sqrt{c}}}{X_{1}^{d}X_{2}^{d}}
+ \tilde{W}({\frac{X_{1}X_{2}}{c^{1/d}}},X_{3},X_{4},X_{5}),
\end{equation}
where $c$ is a one of the parameters
which characterize the complex structures.
After suitable reparametrization,
one obtains the desirable form of \cite{w}
which can represent the embedding of 
the Seiberg-Witten curve
in terms of the ALE space.
If the $K3$ fibre admits the elliptic fibration,
we have more $(-1,1)$ state written in the form
$\keta {j^{- \frac{2d}{n}}}$
with the integer $n > 2$.

%If we have more $(-1,1)$ states
%$\keta {j^{-l}}$
%with
%\begin{equation}
%(j^{-{d/2}})^{2} = j^{-d},
%\end{equation}
%we can see that 
%the $K3$ fibre admits
%the elliptic fibration.

Let us remind that our orbifold example
$W_{2}$ of (\ref{w2})
gives us a new set of relations of 
twist orerators,
or equivalently a new set of Mori vectors.
This fact suggests the interesting possibility
that after the appropriate orbifoldization,
we will get certain \cym s
in which the method of
geometric engineering proposed in \cite{kmv}
will adequately work.

As a simple and interesting example,
consider the superpotential
\begin{equation}
\label{w}
W = 
X_{1}^{12}+X_{2}^{12}+X_{3}^{6}+X_{4}^{6}+X_{5}^{2}.
\end{equation}
This model corresponds to the \cy\ manifold
whose hypersurface is embedded in  
$WCP_{(1,1,2,2,6)}[12]^{2}_{-252}$.
This hypersurface admits $K3$-fibration.
This model contains two $(-1,1)$ states
$\keta {j^{-2}}$ and $\keta {j^{-6}}$.
Their toric data are $(0,0,0,0)$
and $(0,-1,-1,-3)$, respectively.
The dual polyhedron $\Delta^{*}_{W}$
consists of the following vertices
\begin{eqnarray*}
&\nu^*_1=(-1,-2,-2,-6)&,  \ \
\nu^*_2=(1,0,0,0), \ \
\nu^*_3=(0,1,0,0), \\
&\nu^*_4=(0,0,1,0)&, \ \
\nu^*_5=(0,0,0,1), \ \
\nu^*_6=(0,-1,-1,-3),
\end{eqnarray*}
and we have two Mori vectors
$l^{(1)}_{1}=(0;1,1,0,0,0,-2,0)$ and
$l^{(2)}_{1}=(-6;0,0,1,1,3,1,0)$
(the last entry 0 is unnecessary at present,
but we have added it for later use).
These Mori vectors $l^{(1)}_{1}$ and $l^{(2)}_{1}$
correspond to the two relations of the $U(1)$ twist
operators
$(j^{-6})^{2} = j^{-12}$ and
$j^{-6} = (j^{-1})^{6}$ respectively,
as noted in the previous section.
Due to the these Mori vectors,
the method of geometric engineering in \cite{kmv}
does not work as it stands.
However after the appropriate orbifoldization,
we can apply that method
to obtain a quantum field theory.

Consider the $\ZZ_{2}$ orbifoldization
\begin{equation}
%\label{identifyw7}
X_{2} \sim -X_{2}, \ \ 
X_{5} \sim -X_{5}.
\end{equation}
We then obtain the orbifoldized superpotential
\begin{equation}
\label{wprime}
W^{\prime} = 
Y_{1}^{12}+Y_{2}^{6}+Y_{3}^{6}+Y_{4}^{6}+Y_{2}Y_{5}^{2}.
\end{equation}
The corresponding hypersurface is embedded 
in $WCP_{(1,2,2,2,5)}[12]_{-180}^{4}$.
Note that the corresponding hypersurface does not admit
$K3$-fibration.
But the hypersurface which corresponds to
the following transposed potential
\begin{equation}
\label{tildewprime}
\tilde{W^{\prime}} = 
\tilde{Y_{1}}^{12}+\tilde{Y_{2}}^{6}\tilde{Y_{5}}
+\tilde{Y_{3}}^{6}+\tilde{Y_{4}}^{6}+\tilde{Y_{5}}^{2}
\end{equation}
does admit  $K3$-fibration,
since that hypersurface is embedded in 
$WCP_{(1,1,2,2,6)}[12]$.
The original \cy\ manifold is embedded
in the same weighted projective space.
This is the general property for our orbifoldization
that the transposed hypersurface of the orbifoldized one
is embedded in the same weighted projective space
as the original hypersurface is embedded in.
%This is an important property.
This fact is important.
To obtain a quantum field theory,
we have only to consider
the local mirror symmetry of \cite{kmv}.
So at least one of the hypersurfaces corrresponding to
${W^{\prime}}$ and $\tilde{W^{\prime}}$
should have the $K3$-fibration structure.

At the orbifold point,
we still have the two $(-1,1)$ states
$\keta {j^{-2}}$ and $\keta {j^{-6}}$
so that the dual polyhedron $\Delta^{*}_{W}$
is unchanged.
However once we consider the orbifoldized potential
(\ref{wprime}),
we obtain a new set of operators
$\{ {\tilde{\rho}}_{i} \}$ satisfying
\begin{equation}
%\label{rhoalg2}
{\tilde{\rho_{1}}}^{12}
= {\tilde{\rho_{2}}}^{6}{\tilde{\rho_{5}}}
= {\tilde{\rho_{3}}}^{6} = {\tilde{\rho_{4}}}^{6}
= {\tilde{\rho_{5}}}^{2}
= {\rm identity}. 
\end{equation}
Since the new relation of operators
${({{\tilde{\rho}}_{2}}^{6}{{\tilde{\rho}}_{5}})}^{2}
= {{{\tilde{\rho}}_{2}}^{12}}{{{\tilde{\rho}}_{5}}^{2}}
= {\rm identity}$ holds,
we have a new toric data
$\nu^{*}_{7} = (2,0,0,-1)$
which corresponds to
${{{\tilde{\rho}}_{2}}^{12}}
= {\rm identity}$.
The insertion of $\nu^{*}_7$ into the dual polyhedron
$\Delta^{*}_{W}$ implies the toric blow-up.
As a result, we obtain a new set of Mori vectors
\begin{equation}
l^{(1)}_{2}=(0;1,1,0,0,0,-2,0) = l^{(1)}_{1},
\end {equation}
\begin{equation}
l^{(2)}_{2}=(0;0,-2,0,0,1,0,1),
\end {equation}
\begin{equation}
l^{(3)}_{2}=(-6;0,2,1,1,2,1,-1).
\end {equation}
The orbifold relation is
$l^{(2)}_{1} = l^{(2)}_{2} + l^{(3)}_{2}$.
Using $l^{(1)}_{2}$ and $l^{(2)}_{2}$,
we can obtain a quantum field theory
by the geometric engineering of \cite{kmv}.

\section{Discussions}

In this paper, we have given
the toric interpretation of 
\lg\ models.
This interpretation gives us the insight 
and the useful techniques 
to resolve the twisted sector problem.
The orbifold construction is a powerful tool 
for solving the twisted sector problem.
Using mirror symmetry, 
we can study somewhat large class of 
the K\"{a}hler moduli space.
Our construction shows that
through the orbifold point
the different \lg\ string vacua are connected.
It is discussed
that a relation of twist operators for orbifoldization
have a natural interpretation as a Mori vector.

It is natural to expect
that \lg\ models with $c = 12$ correspond to
\cy\ fourfolds.
Recently, \cy\ fourfolds are studied in the context of F-theory
\cite{mv,bs},
i.e. a F-theory on a \cy\ fourfold is a  dual candidate
for a heterotic string on a \cy\ threefold.
Our results of correspondence between the \lg\ model and toric
geometry can be easily extended to \cy\ fourfolds
and give us some insight for those theories.

In \cite{mayr} it is pointed out that for \cy\ fourfolds
on which F-theory can compactify,
the same difficulties as twisted sector problem
for threefolds may occur
when one considers the non-perturbative superpotential
\cite{witten}.
We hope that our approach for solving the twisted sector problem
will be useful to improve the unexpected situations
of \cy\ fourfolds.

Recently, some attention has been paid to 
the extremal transition \cite{extremal}.
It is interesting to consider the relation
between our orbifold construction and such 
extremal transition.
In the context of F-theory,
the 4-cycle transition is studied.
For example \cite{alg}, 
the Fermat hypersurface in $WCP_{(1,2,2,3,4)}[12]_{74}^{2}$
has the 4-cycle transition 
by which this theory transforms to
the Fermat hypersurface in $WCP_{(1,1,1,1,2)}[6]_{103}^{1}$.
It is easy to see that
this transition can not be considered as our orbifold construction.
%the $\ZZ_{2}$ identification is,
%\begin{equation}
%\end{equation}
So, it is interesting to consider
whether our orbifold construction gives us
a new class of examples as extremal transition.

\vspace{1cm}

{\it Acknowledgements} : 
The author would like to thank H. Itoyama
for valuable discussions
and for careful reading of this manuscript.

%%%%%BIB%%%%%%%%%%%%%%%%%%%%%%%%%%%%%%%%%%%%%%%%%%%%%%%%%%%%%%%%%%%%% 
%%%%%%

\end{document}